\documentstyle[12pt,sprocl,epsf,cite]{article}

\newif\ifpreprint
\preprinttrue

\ifpreprint
\fi

\begin{document}

\title{Proof of Factorization for Exclusive Deep-Inelastic
       Processes
\ifpreprint
   \footnote{Presented at Madrid low $x$ workshop, June 18--21, 1997.}
\fi
      }

\author{John C. Collins$^{\dag}$, L. Frankfurt$^{\ddag}$,
        and M. Strikman$^{\dag}$
        \\*[2mm]
}

\address{
    $^{\dag}$ Penn State University, 104 Davey Lab, \\
    University Park PA 16802, U.S.A.
    \\[2mm]
    $^{\ddag}$ Physics Department, Tel-Aviv University, Tel-Aviv, Israel
}

\date{September 12, 1997}

\maketitle
\abstracts{
    This talk summarized the proof of hard-scattering
    factorization for exclusive deep-inelastic processes, such as
    diffractive meson production.
}

\section{Introduction}
\label{sec:introduction}

One of the interesting features of diffractive vector meson
production is that it gives a novel way of probing parton
densities.
Donnachie and Landshoff \cite{DL} constructed a parton model for
the process; their model is incorporated as an approximation in
some sense in all the later work.
Then Ryskin \cite{Ryskin} showed how to estimate $J/\psi $ production
with the use of the BFKL pomeron and a constituent quark model.
Brodsky et al.\ \cite{BFGMS} showed how to treat the production
of light vector mesons; this work was in the leading logarithm
approximation in $x$. After that Frankfurt, Koepf, and Strikman
\cite{FKS} calculated the process in the leading $\ln Q^{2}$
approximation.

It has since been possible to prove a full factorization theorem,
and it is this work \cite{CFS} that is summarized here.

The proof is to all orders of perturbation theory and encompasses
all logarithmic corrections, so that a systematic discussion of
corrections to the leading-logarithm result is now possible.
The parton densities are off-diagonal generalizations of the
usual parton densities \cite{off-diag-pdfs}.

It is interesting that the proof only applies when the virtual
photon is longitudinally polarized.  Transverse polarization for
the photon implies power suppression, relative to the case of
longitudinal polarization, and is therefore much harder to
discuss with the same level of precision.

A new property, discovered while constructing the proof, is that
the theorem applies to the production of all kinds of meson and
at all $x$, in contrast to the original work
\cite{DL,Ryskin,BFGMS,FKS}, which was for vector mesons at small
$x$.  For the case of longitudinally polarized vector mesons, the
parton densities are the ordinary unpolarized ones.  For
transversely polarized vector mesons, the parton densities are
the quark transversity densities, $\delta q$, while for pseudo-scalar
mesons the parton densities are the quark helicity densities,
$\Delta q$.  Hence the polarized parton densities can be probed in
unpolarized collisions.

\section{Theorem}
\label{sec:theorem}

The theorem is for the process
\begin{equation}
     \gamma ^{*}(q) + p \to  M(q+\Delta ) + p'(p-\Delta )
\label{process}
\end{equation}
at large $Q^{2}$, with $t$ and $x=Q^{2}/2p\cdot q$ fixed.  It asserts that
the amplitude has the form
\begin{eqnarray}
   &&
   \sum _{i,j} \int _{0}^{1}dz  \int d\xi
   f_{i/p}(\xi ,\xi -x;t,\mu ) \,
   H_{ij}(Q^{2}\xi /x,Q^{2},z,\mu )
   \, \phi _{j}(z,\mu )
\nonumber\\
&&
   + \mbox{power-suppressed corrections} ,
\label{factorization}
\end{eqnarray}
where $f$ is an ``off-diagonal parton density''
\cite{off-diag-pdfs}, $\phi $ is the light-front wave function of the
meson, and $H$ is a hard-scattering coefficient, usefully
computable in powers of $\alpha _{s}(Q)$.

\section{Proof}
\label{sec:proof}

The proof follows the usual lines of a proof of factorization for
an inclusive hard process \cite{fact}.  We work in a frame in
which the virtual photon, the proton and the meson have momentum
components (in ordinary $(t,x,y,z)$ coordinates)
\begin{eqnarray}
    q^{\mu } &=& (0,0,0,-Q),
\nonumber\\
    p^{\mu } &\approx & (Q/2x, 0, 0, Q/2x),
\nonumber\\
    V^{\mu } &\approx & (Q/2, 0, 0, -Q/2).
\label{components}
\end{eqnarray}
The approximations in the last two lines involve the neglect of
masses and small transverse momenta.

\begin{figure}
    \begin{center}
        \leavevmode
        \epsfxsize=.35\hsize
        \epsfbox{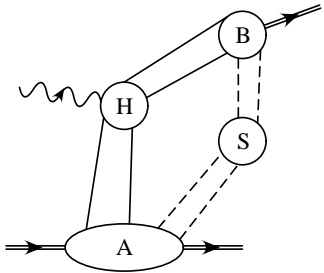}
    \end{center}
\caption{Typical leading region.}
\label{fig:Reduced.Graph}
\end{figure}

Now the usual technology for obtaining the leading power behavior
tells us that this comes from regions symbolized by Fig.\
\ref{fig:Reduced.Graph}, which has groups of lines that are: collinear to
the target proton, collinear to the meson, and hard.  In addition
there may be soft lines joined to the two collinear subgraphs by
gluons.

The primary difficulty is to show that the effects of the soft
gluons cancel.  We consider the attachments of the soft gluons to
the final-state meson.  After making a leading power
approximation suitable to the final-state lines, a Ward identity
can be applied to show that the soft attachments to the meson
subgraph sum to eikonal line connections to the lines coupling
the meson to the hard subgraph.  At this point, it is essential
that the collinear interactions making the meson are all in the
final state relative to the hard scattering.  This enables a
contour deformation to be made, just as in the case \cite{ColSt}
of inclusive $e^{+}e^{-}$ annihilation.  Only after the contour
deformation can the approximation be made that allows Ward
identities to be used.

After that, the color singlet nature of the meson and the
relative point-like nature of the hard scattering are used to
show that the soft interactions cancel.  (We presented this
argument differently in our paper \cite{CFS}.)
The factorization theorem Eq.\ (\ref{factorization}) then follows
easily.

The proof of cancellation of the soft gluon interactions is
intimately related to the fact that the meson arises from a
quark-antiquark pair generated by the hard scattering.  Thus the
pair starts as a small-size configuration and only substantially
later grows to a normal hadronic size, in the meson.  This
implies that the parton density is a standard parton density
(apart from the off-diagonal nature of its definition).  For
example, no rescattering corrections are needed on a nuclear
target, other than those that are implicit in the definition of
universal parton densities, and that would equally appear in
ordinary inclusive deep-inelastic scattering. These statements
all apply to the leading power.

\section{Which parton densities go with which meson?}
\label{sec:which}

\begin{figure}
    \begin{center}
        \leavevmode
        \epsfxsize=0.2\hsize
        \epsfbox{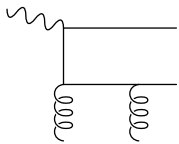}
    \end{center}
\caption{A typical graph for a gluon-induced hard scattering.}
\label{fig:H.graph}
\end{figure}

The general structure of the proof merely shows that the parton
densities in Eq.\ (\ref{factorization}) are any of the usual
leading twist parton densities.  A more detailed argument
involving the spin structure is needed to show which of the
parton densities is needed.  Consider first the Born graphs, such
as Fig.\ \ref{fig:H.graph}, for the hard scattering $H$.

To leading power, we ignore masses in $H$.  Then Fig.\
\ref{fig:H.graph} contains an odd number of Dirac matrices, which
are to be contracted with a matrix from the light-cone meson wave
function.  We have
\begin{equation}
   {\rm tr} \left[
      (\mbox{Odd \# of Dirac matrices})
      \times
      \gamma ^{+} \times  \left(
            \begin{array}{c}
                1  \\
                \gamma _{5} \\
                \gamma _{i}
            \end{array}
         \right)
   \right] ,
\end{equation}
where the column lists the cases for a longitudinally polarized
vector meson, a pseudo-scalar meson, and a transversely polarized
vector meson.  (The index $i$ is a transverse index.)

It follows immediately that we get a zero trace for the case of a
transverse vector meson, which can therefore not be generated by
a gluon-induced process.

In the other two cases, the trace is zero unless the photon is
longitudinal.  Moreover, charge-conjugation invariance kills the
case of the pseudo-scalar meson.  Thus we find that the
gluon-induced subprocess is associated only with the unpolarized
gluon density and only with longitudinal vector meson production
from a longitudinal photon.  All other cases are power
suppressed.

Similar arguments for quark-induced processes give the same
results, except that transverse vector meson production is
associated with the transversity density $\delta q$ and pseudo-scalar
meson production with the helicity density $\Delta q$.

The arguments generalize to all orders of perturbation theory in
the hard scattering.

\begin{figure}
    \begin{center}
        \leavevmode
        \epsfxsize=0.34\hsize
        \epsfbox{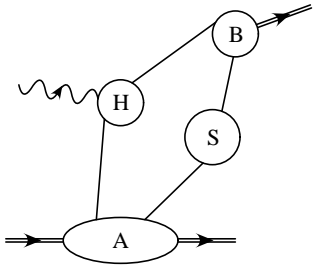}
    \end{center}
\caption{This region is leading for a transversely polarized
         vector meson, provided that there is a sufficiently soft
         quark (with possible gluons).}
\label{fig:endpoint}
\end{figure}

{\it There is one complication}: A transversely polarized vector
meson can result from a hard scattering with just two quark lines
instead of four.  The soft subgraph then has two external quark
lines (plus optional additional gluons) -- Fig.\
\ref{fig:endpoint}.  We gain a power of $Q$ because the hard
scattering has fewer external lines, but lose a power because of
the soft quarks.  This contribution is called an endpoint
contribution, because it probes the meson's wave function at $z
\to 0$ (or 1).  In the case of longitudinal polarization, the
endpoint term is suppressed relative to the non-endpoint term,
but for transverse polarization, both contributions are
comparable until one appeals to a Sudakov suppression.  This
means that it is substantially harder to give a quantitative
discussion of transverse polarization than is the case for
longitudinal polarization.  The most important fact, however, is
that both contributions for transverse polarization are a power
of $Q$ smaller than the amplitude for longitudinal polarization.

\section{Pseudo-scalar meson production}
\label{sec:pion}

Exclusive pion production involves the helicity parton densities.
So it should not be suppressed at large $x$ compared to vector
meson production.  But it should be much smaller at small $x$.

A number of predictions \cite{CFS} can be made for ratios of
cross sections of different mesons, if some approximations are
made.  These are that the meson wave functions are SU(3)
symmetric, that the strange quark helicity density $\Delta s$ is small,
and that the helicity distribution of the up and down quarks are
approximately equal and opposite: $\Delta d\approx -\Delta u$ (this follows
from the
observation that $F_{2}$ for the deuteron is small and the
assumption that this same property is valid for the off-diagonal
parton densities).  We therefore predict that
\begin{eqnarray}
    \frac {d\sigma (e+p \to  \eta +p)/dt}{d\sigma (e+p \to  \pi ^{0}+p)/dt}
    &\approx &
    \frac {1}{3} \left(
         \frac {2\Delta u_{V} - \Delta d_{V}}{2\Delta u_{V} + \Delta d_{V}}
      \right)^{2}
    \approx  3,
\nonumber\\
    \frac {d\sigma (e+p \to  \eta +p)/dt}{d\sigma (e+n \to  \eta +n)/dt}
    &\approx &
      \left(
         \frac {2\Delta u_{V} - \Delta d_{V}}{2\Delta d_{V} - \Delta u_{V}}
      \right)^{2}
    \approx  1,
\nonumber\\
    \frac {d\sigma (e+p \to  \pi ^{0}+p)/dt}{d\sigma (e+n \to  \pi ^{0}+n)/dt}
    &\approx &
      \left(
         \frac {2\Delta u_{V} + \Delta d_{V}}{2\Delta d_{V} + \Delta u_{V}}
      \right)^{2}
    \approx  1.
\end{eqnarray}
Here $\Delta u_{V}=\Delta u-\Delta \bar u$ and $\Delta d_{V}=\Delta d-\Delta
\bar d$.

\section{Conclusions}
\label{sec:conclusions}

We have a full proof of hard-scattering factorization not only
for exclusive vector meson production in DIS, but also for
processes such as $\gamma ^{*} + p \to  \pi ^{+} + n$.  Among the new results
are
that: (a) the theorem holds for large $x$ as well as small $x$,
(b) it works for any meson, (c) some polarized parton densities,
including the elusive transversity density, can be probed in
unpolarized collisions, and (d) there is a power suppression of
transverse vector meson production, beyond leading-logarithm
approximation, but only at small $x$, because of the properties
of polarized parton densities.

\section*{Acknowledgments}

This work was supported in part by the U.S.\ Department of Energy
under grant number DE-FG02-90ER-40577 and DE-FG02-93ER40771, and
by the Binational Science Foundation under Grant No.\ 9200126.



\begin{thebibliography}{99}

\bibitem{DL}
    A. Donnachie and P.V. Landshoff, {\em Phys.\ Lett.\ B} {\bf 185},
    403 (1987); {\em Nucl.\ Phys.\ B} {\bf 311}, 509 (1989).

\bibitem{Ryskin}
   M.G. Ryskin, {\em Z. Phys.\ C} {\bf 57}, 89 (1993).

\bibitem{BFGMS}
   S.J. Brodsky, L. Frankfurt, J.F. Gunion, A.H. Mueller, and M.
   Strikman, {\em Phys.\ Rev.\ D} {\bf 50}, 3134 (1994).

\bibitem{FKS}
   L. Frankfurt, W. Koepf, and M. Strikman, Phys.\ Rev.\ D
   {\bf 54}, 3194 (1996).

\bibitem{CFS}
    J.C. Collins, L. Frankfurt, and M. Strikman,
    {\em Phys.\ Rev.\ D} {\bf 56}, 2982 (1997).

\bibitem{off-diag-pdfs}
    J. Bartels and M. Loewe, {\em Z. Phys.\ C} {\bf 12}, 263 (1982);
    \\
    B. Geyer {\em et al.}, {\em Z. Phys.\ C} {\bf 26}, 591 (1985);
    \\
    T. Braunschweig {\em et al.}, {\em Z. Phys.\ C} {\bf 33}, 275 (1987);
    \\
    I.I. Balitsky and V.M. Braun, {\em Nucl.\ Phys.\ B} {\bf 311}, 541
    (1989);
    \\
    A.V. Radyushkin, {\em Phys.\ Lett.\ B} {\bf 380}, 417 (1996);
    \\
    X.-D. Ji, {\em Phys.\ Rev.\ D} {\bf 55}, 7114 (1997).

\bibitem{fact}
   J.C. Collins, D.E. Soper, and G. Sterman, {\em Nucl.\ Phys.\
   B} {\bf 261}, 104 (1985) and {\bf B308}, 833 (1988);
   \\
   G. Bodwin, {\em Phys.\ Rev.\ D} {\bf 31}, 2616 (1985); {\bf 34},
   3932 (1986).

\bibitem{ColSt}
   J.C. Collins and G. Sterman, {\em Nucl.\ Phys.\ B} {\bf 185}, 172 (1981).



\end{thebibliography}
\end{document}